\documentclass[pdflatex,sn-mathphys-num]{sn-jnl}% Math and Physical Sciences Numbered Reference Style
\usepackage{graphicx}%
\usepackage{multirow}%
\usepackage{amsmath,amssymb,amsfonts}%
\usepackage{amsthm}%
\usepackage{mathrsfs}%
\usepackage[title]{appendix}%
\usepackage{xcolor}%
\usepackage{textcomp}%
\usepackage{manyfoot}%
\usepackage{booktabs}%
\usepackage{algorithm}%
\usepackage{algorithmicx}%
\usepackage{algpseudocode}%
\usepackage{listings}%
\theoremstyle{thmstyleone}%
\theoremstyle{thmstyletwo}%

\theoremstyle{thmstylethree}%

\begin{document}

\title[Article Title]{LPGNet: A Lightweight Network with Parallel Attention and Gated Fusion for Multimodal Emotion Recognition}


\author[1]{\fnm{Zhining} \sur{He}}\email{hezhining0624@163.com}

\author*[2]{\fnm{Yang} \sur{Xiao}}\email{yxiao9550@student.unimelb.edu.au}

\affil[1]{\orgname{Guangzhou University}, \orgaddress{\city{Guangzhou}, \country{China}}}

\affil*[2]{\orgname{The Unviersity of Melbourne}, \orgaddress{\city{Melbourne}, \country{Australia}}}


\abstract{Emotion recognition in conversations (ERC) aims to predict the emotional state of each utterance by using multiple input types, such as text and audio. While Transformer-based models have shown strong performance in this task, they often face two major issues: high computational cost and heavy dependence on speaker information. These problems reduce their ability to generalize in real-world conversations. To solve these challenges, we propose LPGNet, a Lightweight network with Parallel attention and Gated fusion for multimodal ERC. The main part of LPGNet is the Lightweight Parallel Interaction Attention (LPIA) module. This module replaces traditional stacked Transformer layers with parallel dot-product attention, which can model both within-modality and between-modality relationships more efficiently. To improve emotional feature learning, LPGNet also uses a dual-gated fusion method. This method filters and combines features from different input types in a flexible and dynamic way. In addition, LPGNet removes speaker embeddings completely, which allows the model to work independently of speaker identity. Experiments on the IEMOCAP dataset show that LPGNet reaches over 87\% accuracy and F1-score in 4-class emotion classification. It outperforms strong baseline models while using fewer parameters and showing better generalization across speakers. }




\maketitle

\section{Introduction}\label{sec1}

Emotion recognition in conversations (ERC)~\cite{poria2019emotion,qin2023bert} focuses on identifying the emotional state of each utterance by using multiple types of input, such as text, audio, and visual signals. It plays an important role in human-computer interaction (HCI)~\cite{ramakrishnan2013speech,cowie2001emotion}, especially in building systems that can understand and respond to users’ emotions. ERC is a key part of many real-world applications, including empathetic dialogue systems, emotion-aware virtual assistants, and mental health support tools. These systems are widely used in areas like intelligent voice assistants, customer service robots, and telemedicine. Compared to traditional emotion recognition based on a single input type, ERC brings more challenges. This is because it requires understanding context over time, handling frequent speaker changes, and processing emotional signals across different modalities. These challenges make ERC a complex but important problem in affective computing.

Recent ERC models based on Transformer~\cite{ma2023transformer} and graph neural network (GNN)~\cite{ghosal2019dialoguegcn,chen2023multivariate,song2023sunet} architectures have shown strong performance. However, they often face two key challenges. First, deep sequential layers lead to high computational cost, which limits deployment in real-time systems. Second, many models depend heavily on speaker identity features, making them less effective in open-domain or anonymous settings, where speaker information is missing or unavailable. These limitations highlight the need for efficient and speaker-independent ERC models.

To improve multimodal fusion in ERC, several attention-based approaches have been developed. These methods aim to better integrate features from different input types, such as text and audio. For example, Tang et al.~\cite{tang2022bimodal} introduced an audio-text interactive attention mechanism that strengthens cross-modal understanding. Zhao et al.~\cite{zhao2023bayesian} proposed collaborative attention to align emotional information across pre-trained features. Maji et al.~\cite{maji2023multimodal} designed a cross-modal Transformer block to capture both semantic and temporal dependencies across modalities. Although these techniques have demonstrated good performance, they often come with a high cost. Their architectures are usually deep and complex, which increases both training and inference time. As a result, these models are not well-suited for real-time or resource-constrained environments, such as mobile devices.

In addition to attention-based models, graph neural networks (GNNs) have also been applied to ERC. Methods such as DialogueGCN~\cite{ghosal2019dialoguegcn} and MMGCN~\cite{hu2021mmgcn} aim to capture speaker interactions by constructing dialogue graphs. These models can represent complex conversational structures but rely heavily on speaker identity labels and manually crafted graphs. This dependency reduces their flexibility and limits their usefulness in open-domain applications, where such information may be unavailable or inconsistent.

These limitations become especially important when considering real-world ERC applications. In practical settings, systems must often deal with unknown, anonymous, or changing speakers~\cite{cl1,cl2,cl3}. They also need to perform emotion recognition in a zero-shot manner, without prior training on speaker-specific data. For instance, voice assistants must understand a user's emotions during their first interaction, without relying on stored speaker profiles~\cite{ssl1}. Similarly, customer service bots interact with a wide range of users, many of whom are anonymous or change frequently. In such cases, speaker embeddings offer little benefit. Multi-user devices, such as smart home terminals, require models to handle frequent speaker switching without identity tracking.

Recent ERC methods based on large language models (LLMs)~\cite{llm1,llm2,llm3} also reveal inconsistencies in how speaker information is handled. InstructERC~\cite{wu2024beyond} treats ERC as a generation task using emotion templates, but it does not explicitly model speaker roles. DialogueLLM~\cite{zhang2025dialoguellm} uses dialogue context and visual cues but omits speaker identity. BiosERC~\cite{xue2024bioserc} goes further by showing that privacy constraints often prevent the use of identity labels, meaning that models must rely only on the dialogue context. These findings underline the importance of speaker-independent emotion modeling for real-world applications.

Alongside attention and graph-based techniques, self-distillation~\cite{zhang2019your,zhang2021self,xiao2024ucil} has emerged as a promising solution for improving model performance without increasing complexity. Unlike traditional distillation methods that rely on an external teacher model, self-distillation enables a network to learn from its own intermediate features. For example, Li et al.~\cite{li2022teacher} introduced a teacher-free feature distillation method in vision tasks, where a model reuses useful internal representations to boost its own learning. This strategy reduces both memory and computation costs, making it well-suited for lightweight deployment.

Based on these insights, we propose LPGNet, a lightweight and speaker-independent framework for multimodal emotion recognition in conversations. LPGNet is designed to address three major issues: high inference cost, reliance on speaker information, and the need for efficient deployment in practical scenarios.

Our key contributions are as follows:

\begin{itemize}
  \item We design a Lightweight Parallel Interaction Attention (LPIA) module to replace stacked Transformer layers, allowing efficient modeling of intra- and inter-modal relationships.
  \item We introduce a dual-gated fusion strategy to refine and dynamically combine multimodal features.
  \item We remove speaker embeddings entirely, enabling better generalization to unseen speakers and anonymous environments.
  \item We apply a self-distillation mechanism to internally extract and reuse useful emotional knowledge without requiring external teacher models.
\end{itemize}

Extensive experiments on the IEMOCAP benchmark show that LPGNet achieves over 87\% accuracy and F1-score for 4-class emotion classification. It not only outperforms strong baseline models but also shows better generalization to unknown speakers, all while using fewer parameters.

\section{Related Work}
\subsection{Multimodal Fusion Strategies in Emotion Recognition}

Multimodal emotion recognition (MER)~\cite{zhang2024deep} aims to integrate signals from different modalities such as text, speech, and vision. Early models relied on recurrent neural networks and memory-based fusion techniques. For example, CMN~\cite{hazarika2018conversational} used GRUs to model speaker history and applied attention to extract relevant emotion patterns from past utterances. ICON~\cite{hazarika2018icon} extended this by introducing interactive memory modules to model affective influence between speakers. These methods successfully leveraged dialogue context but involved complex architectures, making them difficult to scale for large or real-time applications.

More recently, Transformer-based methods have become dominant. MER-HAN~\cite{zhang2023multimodal} uses a three-stage attention framework to model intra-modal, cross-modal, and global features. While this improves interpretability, it increases model complexity and may suffer from misalignment across modalities. To improve efficiency, models like SDT~\cite{ma2023transformer} apply parallel intra- and cross-modal attention combined with gated fusion and self-distillation. XMBT~\cite{nguyen2025enhanced} introduces shared bottleneck tokens for cross-modal interaction, which helps reduce inference cost and allows flexible modality integration.

Despite performance gains, these methods still face key limitations. Many use deep sequential Transformers that are computationally expensive. Others lack dynamic weighting, treating all modalities equally—even when one modality carries more emotional information than others. To address this, LPGNet introduces a Lightweight Parallel Interaction Attention (LPIA) module to replace stacked layers and reduce computation. It also includes a dual-gated fusion mechanism to dynamically filter and combine multimodal features based on signal strength and context.

\subsection{Speaker Dependency and Real-World Limitations}

Several advanced MER models, including COSMIC~\cite{ghosal2020cosmic} and DialogXL~\cite{shen2021dialogxl}, incorporate speaker identity information or model speaker-role interactions through GNNs. While this improves performance in speaker-labeled datasets, it reduces flexibility in real-world conditions where speaker labels are unavailable or privacy constraints exist. For instance, voice assistants and multi-user devices often encounter new or anonymous speakers. In such cases, models depending on speaker embeddings may fail to generalize. Recent research has explored speaker-independent modeling. SIMR~\cite{kuhlen2022mental} attempts to remove identity bias by disentangling style and semantic content in non-verbal modalities. However, most systems still include speaker-level features or require predefined speaker roles, limiting their adaptability in open-domain dialogue.

Existing approaches do not fully eliminate speaker dependence. In contrast, LPGNet is designed to be speaker-independent by default. It removes all speaker-related inputs and structures, allowing the model to generalize across unknown speakers and dynamically changing roles—an essential requirement for scalable and privacy-aware systems.

\subsection{Lightweight Learning with Self-Distillation}

As emotion recognition moves toward real-time and mobile applications, efficiency has become a core concern. Traditional models like MER-HAN and SDT require deep Transformer stacks, which increase latency and memory usage. Knowledge distillation (KD) has been proposed to compress large models while maintaining accuracy. For example, DMD~\cite{yin2022mix} uses cross-modal distillation to decompose and transfer shared and unique features. MT-PKDOT~\cite{aslam2024multi} applies multi-teacher distillation to guide student models using diverse sources. While effective, these strategies often require additional teacher models and increase training complexity. Some models, like SDT, use self-distillation to reuse internal knowledge, but they still rely on heavy architectures.

Few models combine self-distillation with lightweight structures. LPGNet adopts a teacher-free self-distillation approach within a compact design. It transfers knowledge across internal layers without needing extra networks, improving performance and efficiency. This allows LPGNet to remain accurate while significantly lowering inference cost and parameter size, making it suitable for real-world, resource-constrained deployment.

\section{Proposed Method}

\subsection{Task Definition}
The input to the ERC task is a conversation composed of $N$ utterances $\{u_1, u_2, \dots, u_N\}$, where each utterance $u_i$ contains both textual and acoustic modalities. After feature extraction, the text and audio features are projected to a common latent space:

\begin{equation}
\mathbf{X}_t \in \mathbb{R}^{B \times U_i \times F}, \quad \mathbf{X}_a \in \mathbb{R}^{B \times U_i \times F}
\end{equation}

Here, $B$ is the batch size, $U_i$ denotes the number of utterances, and $F$ is the feature dimension. The goal of ERC is to predict an emotion label for each utterance $u_i$ from a set of predefined emotion classes.

\subsection{Overview}
\begin{figure}
    \centering
    \includegraphics[width=1\linewidth]{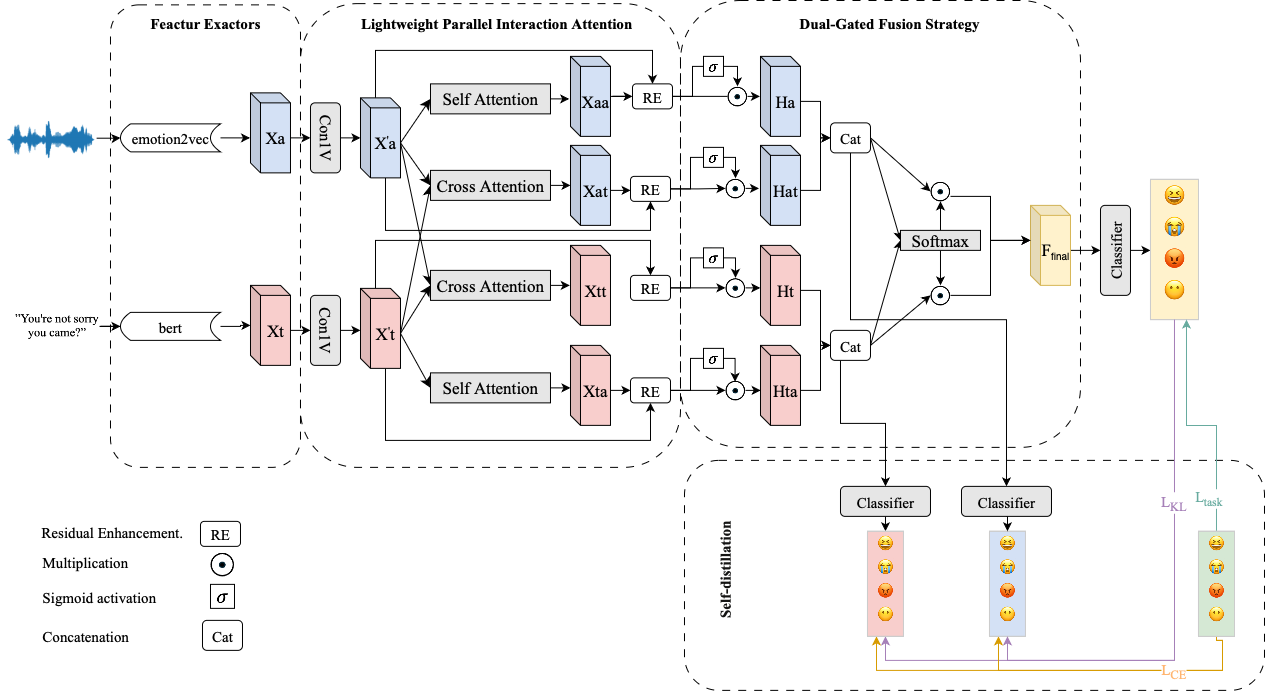}
    \caption{Overall architecture of LPGNet. After extracting utterance-level unimodal features, it consists of four key components: LPIA Block, Dual Gated Fusion, Classifier, and Self-distillation}
    \label{fig:LPGNet}
\end{figure}

Figure~\ref{fig:LPGNet} provides an overview of the proposed \textbf{LPGNet} model. After extracting utterance-level acoustic and textual features, LPGNet consists of four major modules: 
\begin{itemize}
    \item \textbf{Lightweight Parallel Interaction Attention (LPIA)} module for capturing both intra- and inter-modal interactions simultaneously;
    \item \textbf{Dual-Gated Fusion} module for refining unimodal representations and adaptively fusing multimodal information;
    \item \textbf{Emotion Classifier} that predicts emotion labels from the final fused representation;
    \item \textbf{Self-Distillation mechanism} that aligns unimodal branches with the fused multimodal output using both hard and soft label supervision.
\end{itemize}

% Figure~\ref{fig2}

\subsection{Lightweight Parallel Interaction Attention (LPIA)}

To effectively model both intra- and inter-modal dependencies in a lightweight manner, we propose a parallel attention structure named \textbf{Lightweight Parallel Interaction Attention (LPIA)}. This module aims to fully leverage the complementary characteristics of acoustic and textual modalities through parallel attention mechanisms.

\paragraph{(1) Modality-wise Local Encoding via 1D Convolution.}
We first apply a 1D convolution with kernel size 1 to both textual and acoustic input sequences. This operation serves two purposes: (1) to project modality-specific features into a unified latent space of dimension $d$; and (2) to enable localized context modeling across adjacent utterances within each modality:
\begin{equation}
\mathbf{X}_s' = \text{Conv1D}(\mathbf{X}_s), \quad s \in \{t, a\}
\end{equation}
where $\mathbf{X}_s \in \mathbb{R}^{B \times U \times F_s}$ denotes the original feature sequence for modality $s$ (text or audio), $B$ is the batch size, $U$ is the number of utterances, and $F_s$ is the input feature dimension.

\paragraph{(2) Four-Way Parallel Attention Modeling.}
To simultaneously capture intra- and inter-modal relationships, we define four parallel attention blocks based on scaled dot-product attention:
\begin{equation}
\text{Attention}(Q, K, V) = \text{Softmax}\left(\frac{QK^\top}{\sqrt{d}}\right)V
\end{equation}
The four interaction paths are:
\begin{itemize}
    \item $\mathbf{X}_{tt} = \text{Attention}(\mathbf{X}_t', \mathbf{X}_t', \mathbf{X}_t')$: intra-text attention for modeling semantic continuity;
    \item $\mathbf{X}_{aa} = \text{Attention}(\mathbf{X}_a', \mathbf{X}_a', \mathbf{X}_a')$: intra-acoustic attention for modeling prosodic dynamics;
    \item $\mathbf{X}_{at} = \text{Attention}(\mathbf{X}_a', \mathbf{X}_t', \mathbf{X}_t')$: cross-modal audio-to-text interaction;
    \item $\mathbf{X}_{ta} = \text{Attention}(\mathbf{X}_t', \mathbf{X}_a', \mathbf{X}_a')$: cross-modal text-to-audio interaction.
\end{itemize}
This fully parallel structure avoids sequential stacking, reduces inference latency, and allows efficient modeling of bidirectional modality interactions.

\paragraph{(3) Mask-Aware Attention Computation.}
To ensure robustness in variable-length conversations, we adopt a mask-aware attention mechanism. Let $\mathbf{M} \in \{0,1\}^{B \times U}$ denote a binary mask indicating valid utterances. It is expanded to $\mathbf{M}' \in \{0,1\}^{B \times U \times U}$ for masking attention scores:
\begin{equation}
\mathbf{S}_{i,j}^{(b)} =
\begin{cases}
\frac{\mathbf{Q}_i^{(b)} \cdot \mathbf{K}_j^{(b)}}{\sqrt{d}}, & \text{if } \mathbf{M}_{i,j}^{(b)} = 1 \\
-10^9, & \text{otherwise}
\end{cases}
\end{equation}
The masked scores are normalized by softmax:
\begin{equation}
\mathbf{A}^{(b)} = \text{Softmax}(\mathbf{S}^{(b)})
\end{equation}
This formulation ensures that attention is only paid to valid utterances, suppressing the influence of padding noise.

\paragraph{(4) Two-Stage Residual Enhancement.}
After computing the attention output $\hat{\mathbf{A}} \in \mathbb{R}^{B \times U \times d}$, we apply two refinement stages:
\begin{itemize}
    \item \textbf{1. Global Compression:} We first compress the global context using a lightweight convolutional block composed of 1×1 convolution, batch normalization, and LeakyReLU activation:
    \begin{equation}
    \mathbf{A}_{\text{conv}} = \text{Conv2D}(\text{Mean}(\hat{\mathbf{A}}))
    \end{equation}
    
    \item \textbf{2. Position-wise FFN with Residual Fusion:} The compressed feature is broadcast back to the sequence dimension and fused with the original query and a position-wise feedforward network:
    \begin{equation}
    \text{FFN}(x) = \mathbf{W}_2 \left( \text{Dropout}_1 \left( \text{GELU}(\mathbf{W}_1 \cdot \text{LayerNorm}(x)) \right) \right)
    \end{equation}
    \begin{equation}
    \mathbf{X}_{\text{final}} = \mathbf{Q} + \mathbf{A}_{\text{conv}} + \text{FFN}(\mathbf{A}_{\text{conv}})
    \end{equation}

Here, \( \mathbf{W}_1 \in \mathbb{R}^{d \times d_{ff}} \) and \( \mathbf{W}_2 \in \mathbb{R}^{d_{ff} \times d} \) are learnable weight matrices, where \( d \) is the hidden dimension of the model and \( d_{ff} \) is the intermediate feedforward dimension (typically \( 2d \) or \( 4d \)). The GELU function introduces smooth non-linearity, and Dropout is used for regularization.
\end{itemize}

This two-stage refinement integrates both global context and local non-linearity, enhancing feature expressiveness and gradient stability.

%%%%%%%%%%%%%%%%%%%%%%%%%%%%%%%%%%%%%%%%%%
 \subsection{Dual-Gated Fusion Strategy}

To enable both fine-grained intra-modal enhancement and flexible cross-modal integration, we introduce a hierarchical \textbf{Dual-Gated Fusion} mechanism, composed of two sequential stages: \textit{Unimodal Gated Fusion} and \textit{Multimodal Gated Fusion}.

\paragraph{(1)Unimodal Gated Fusion}

Given the outputs from the intra-modal and inter-modal attention modules, we apply a learnable gating operation to each stream independently. Let $\mathbf{H} \in \mathbb{R}^{B \times U \times d}$ be the hidden representation of an attention output. The refined representation is computed as:
\begin{equation}
\mathbf{Z} = \sigma(\mathbf{W}_g \mathbf{H}), \quad \mathbf{H}_{\text{gated}} = \mathbf{Z} \odot \mathbf{H}
\end{equation}
where $\mathbf{W}_g \in \mathbb{R}^{d \times d}$ is a learnable weight matrix and $\sigma(\cdot)$ denotes the sigmoid activation function. This mechanism allows the model to emphasize informative features while suppressing irrelevant components.

For each modality, we gate the intra-modal and cross-modal branches separately:
\begin{equation}
\mathbf{H}_t = \text{GatedFusion}(\mathbf{X}_{output\_tt}), \quad \mathbf{H}_{at} = \text{GatedFusion}(\mathbf{X}_{output\_at})
\end{equation}
\begin{equation}
\mathbf{H}_a = \text{GatedFusion}(\mathbf{X}_{output\_aa}), \quad \mathbf{H}_{ta} = \text{GatedFusion}(\mathbf{X}_{output\_ta})
\end{equation}

Then, we concatenate and linearly project each modality-specific pair:
\begin{equation}
\mathbf{T}_{\text{fused}} = \text{Linear}([\mathbf{H}_t; \mathbf{H}_{at}]), \quad
\mathbf{A}_{\text{fused}} = \text{Linear}([\mathbf{H}_a; \mathbf{H}_{ta}])
\end{equation}

\paragraph{(2)Multimodal Gated Fusion}

The refined textual and acoustic features are then passed into a multimodal gating mechanism that adaptively assigns weights to each modality. Given the fused features $\mathbf{T}_{\text{fused}}, \mathbf{A}_{\text{fused}} \in \mathbb{R}^{B \times U \times d}$, the attention weights are computed as:
\begin{equation}
\boldsymbol{\alpha}_T = \text{Softmax}(\mathbf{W} \mathbf{T}_{\text{fused}}), \quad
\boldsymbol{\alpha}_A = \text{Softmax}(\mathbf{W} \mathbf{A}_{\text{fused}})
\end{equation}

Let $\mathcal{F} = [\mathbf{T}_{\text{fused}}, \mathbf{A}_{\text{fused}}] \in \mathbb{R}^{B \times U \times 2 \times d}$, the final multimodal representation is computed as:
\begin{equation}
\mathbf{F}_{\text{final}} = \boldsymbol{\alpha}_T \cdot \mathbf{T}_{\text{fused}} + \boldsymbol{\alpha}_A \cdot \mathbf{A}_{\text{fused}}
\end{equation}

This two-stage gating architecture enables the model to dynamically refine unimodal streams and adaptively integrate multimodal context based on conversational cues.

%%%%%%%%%%%%%%%%%%%%%%%%%%%%%%%%%%%%%%%%%%

\subsection{Emotion Classifier}

To obtain the emotion distribution over $C$ categories, the final multimodal representation $\mathbf{F}_{\text{final}} \in \mathbb{R}^{N \times d}$ is passed through a fully connected layer followed by a softmax activation:

\begin{equation}
\mathbf{E} = \mathbf{W}_c \cdot \mathbf{F}_{\text{final}} + \mathbf{b}_c \in \mathbb{R}^{N \times C}
\end{equation}
\begin{equation}
\hat{\mathbf{Y}} = \text{softmax}(\mathbf{E})
\end{equation}

Here, $\mathbf{W}_c \in \mathbb{R}^{d \times C}$ and $\mathbf{b}_c \in \mathbb{R}^C$ are trainable parameters. Let $\hat{\mathbf{Y}} = [\hat{\mathbf{y}}_1; \hat{\mathbf{y}}_2; \ldots; \hat{\mathbf{y}}_N]$, where $\hat{\mathbf{y}}_i$ denotes the predicted emotion probability vector for utterance $u_i$. The predicted label is determined by $\arg\max(\hat{\mathbf{y}}_i)$.

The task loss is defined using the standard cross-entropy objective:

\begin{equation}
\mathcal{L}_{\text{task}} = -\frac{1}{N} \sum_{i=1}^{N} \sum_{j=1}^{C} y_{i,j} \log \hat{y}_{i,j}
\end{equation}

%%%%%%%%%%%%%%%%%%%%%%%%%%%%%%%%%%%%%%%%%%
 \subsection{Self-Distillation}

To further improve modal expressiveness, we adopt a self-distillation strategy that transfers knowledge from the full model (teacher) to each unimodal branch (student). Specifically, we treat the complete model as the teacher and introduce two student branches—textual ($t$) and acoustic ($a$)—trained only during training.

For each modality $m \in \{t, a\}$, the student predicts the emotion distribution via:

\begin{equation}
\mathbf{E}^m = \mathbf{W}_m' \cdot \text{ReLU}(\mathbf{H}'_m) + \mathbf{b}_m' \in \mathbb{R}^{N \times C}
\end{equation}
\begin{equation}
\hat{\mathbf{Y}}^m = \text{softmax}(\mathbf{E}^m), \quad
\hat{\mathbf{Y}}_\tau^m = \text{softmax}(\mathbf{E}^m / \tau)
\end{equation}

where $\tau$ is the temperature for softening the output distribution. Higher values of $\tau$ yield softer class probabilities. $\mathbf{W}_m' \in \mathbb{R}^{d \times C}$ and $\mathbf{b}_m' \in \mathbb{R}^{C}$ are learnable parameters.

Each student is optimized with two losses: The first one is the Cross-Entropy Loss (hard labels) as:
\begin{equation}
\mathcal{L}_{\text{CE}}^m = -\frac{1}{N} \sum_{i=1}^{N} \sum_{j=1}^{C} y_{i,j} \log \hat{y}^m_{i,j}
\end{equation}

And the soft label loss is KL Divergence Loss as:
\begin{equation}
\mathcal{L}_{\text{KL}}^m = \frac{1}{N} \sum_{i=1}^{N} \sum_{j=1}^{C} \hat{y}_{\tau,i,j}^m \log \frac{\hat{y}_{\tau,i,j}^m}{\hat{y}_{\tau,i,j}^{\text{teacher}}}
\end{equation}

Finally, the Total Loss could be formulated as: 
The overall training loss combines task supervision and both hard/soft label distillation:
\begin{equation}
\mathcal{L} = \lambda_{\text{task}} \cdot \mathcal{L}_{\text{task}} 
+ \lambda_{\text{CE}} \cdot \sum_{m} \mathcal{L}_{\text{CE}}^m 
+ \lambda_{\text{KL}} \cdot \sum_{m} \mathcal{L}_{\text{KL}}^m
\end{equation}

\section{Experiments and Results}

%%%%%%%%%%%%%%%%%%%%
\subsection{Dataset and Evaluation}

We evaluate our model using the Interactive Emotional Dyadic Motion Capture (IEMOCAP)~\cite{busso2008iemocap} dataset, a widely adopted benchmark in multimodal emotion recognition. This dataset consists of approximately 12 hours of audio-visual recordings collected from ten professional actors (five male and five female) performing both scripted and improvised dialogues. Each utterance in the dataset is manually annotated with emotional labels across multiple modalities, including text, audio, video, and motion capture data.

In our study, we focus only on the transcribed text and speech signals, which are the most accessible and commonly used modalities in practical applications. Following standard practice in previous works, we define the emotion recognition in conversation (ERC) task as a four-class classification problem. The target emotion categories include \textit{angry}, \textit{happy}, \textit{sad}, and \textit{neutral}. The excited class, which shares similar characteristics with happy, is merged into it to simplify the classification task. After preprocessing and filtering, we obtain a total of 5,531 annotated utterances. The detailed emotion distribution across the dataset is presented in Table~\ref{tab:dataset_distribution}, providing insight into the class balance and evaluation.

\begin{table}[t!]
\centering
\caption{Emotion distribution in the IEMOCAP dataset.}
\label{tab:dataset_distribution}
\begin{tabular}{lcccc|c}
\toprule
Split & Happy & Sad & Neutral & Angry & Total \\
\midrule
Train + Val & 1262 & 828 & 1319 & 872 & 4290 \\
Test        & 442  & 245 & 384  & 170 & 1241 \\
\midrule
Total       & 1704 & 1073 & 1703 & 1042 & 5531 \\
\bottomrule
\end{tabular}
\end{table}

For evaluation, we adopt average binary accuracy and F1-score as our main performance metrics, following the standard protocol used in prior work. Specifically, we compute results using a one-vs-all classification strategy for each emotion class. The final scores are then averaged across all classes. This approach ensures that performance is measured fairly, especially in the presence of class imbalance, by giving equal weight to each class regardless of its frequency in the dataset.

\subsection{Feature Extraction}
\textbf{Text Modality:} To represent the text modality, we use the BERT-base-uncased model as our text encoder. This model consists of 12 Transformer layers, each with 768-dimensional hidden states. We fine-tune BERT on the emotion recognition in conversation task using the dialogue data. For each utterance, we extract the final-layer embedding of the [CLS] token, which serves as a summary representation of the entire sentence.

\textbf{Acoustic Modality:} For the audio modality, we adopt Emotion2vec, a self-supervised acoustic representation model trained on 262 hours of emotional speech data. Emotion2vec is designed to learn both utterance-level and frame-level emotional features through a joint optimization process based on knowledge distillation. From this model, we extract 768-dimensional utterance-level embeddings, which provide rich and generalizable representations of the speech signal.

%%%%%%%%%%%%%%%%%%%%
\subsection{Implementation Details}
Our model is implemented in PyTorch 1.8.1 and RTX 3090. We use the Adam optimizer with a learning rate of 3e-4, a batch size of 32, and a hidden dimension of 768. The temperature parameter $\tau$ for self-distillation is set to 1. We apply L2 weight decay of $1 \times 10^{-5}$ and dropout with a rate of 0.1. Each model is trained for 150 epochs, and we report average results over 10 independent runs for robustness.

%%%%%%%%%%%%%%%%%%%%
\subsection{Results}
Table~\ref{tab:overall_results} presents a detailed comparison between our proposed LPGNet and several state-of-the-art baselines on the IEMOCAP dataset under the 4-class emotion classification setting. Among existing methods, CFIA and MemoCMT show strong performance, with CFIA achieving 83.37\% accuracy using the same BERT and Emotion2vec features as our model. However, our full model, LPGNet (Utterance-level), achieves the highest performance with 87.99\% accuracy and 87.96\% F1-score, representing a relative improvement of over 4\% compared to CFIA. These results highlight the effectiveness of our lightweight and speaker-independent design.

To ensure that this improvement is due to architectural innovation rather than feature quality, we build a minimal baseline using a simple linear classifier applied to the same BERT and Emotion2vec embeddings. This baseline reaches only 81.68\% accuracy, which is 6.31\% lower than our full model. This gap confirms that LPGNet’s performance gains are not solely due to better features, but come from its efficient multimodal fusion and context modeling capabilities.

We further evaluate a frame-level version of our model, called LPGNet (Frame), where both text tokens and acoustic frames are padded or truncated to a fixed length of 512. Despite using finer temporal resolution, LPGNet (Frame) still achieves strong results with 83.87\% accuracy and 83.68\% F1-score, outperforming most baselines. This demonstrates that our model can generalize across different input granularities without losing effectiveness. In summary, LPGNet consistently outperforms existing methods across both utterance- and frame-level settings. These results confirm its robustness, efficiency, and practical potential for real-world multimodal emotion recognition tasks.

\begin{table}[t!]
\centering
\caption{Overall performance comparison on IEMOCAP (4-class).}
\label{tab:overall_results}
\begin{tabular}{lcccc}
\toprule
Model & Upstream(A) & Upstream(T) & ACC (\%) & F1 (\%) \\
\midrule
GatedxLSTM~\cite{li2025gatedxlstm} & CLAP & CLAP & -- & 75.97±1.38 \\
MER-HAN~\cite{zhang2023multimodal} & Bi-LSTM & BERT & 74.20 & 73.66 \\
MMI-MMER~\cite{wang2025emotion} & Wav2Vec2 & BERT &77.02 & -- \\
MGCMA~\cite{wang2025enhancing} & Wav2Vec2 & BERT &78.87 & -- \\
MEP~\cite{shi2023emotion} & openSMILE & BERT & 80.18 & 80.01 \\
Bi-GRU~\cite{adeel2024enhancing} & Acoustic model & GloVe & 80.63 & -- \\
MemoCMT\cite{khan2025memocmt} & HuBERT & BERT & 81.85 & -- \\
Linear\cite{ma2023emotion2vec} & emotion2vec & BERT & 81.68 & 80.75 \\
CFIA\cite{hu2024cross} & emotion2vec & BERT & 83.37 & -- \\
\midrule
\textbf{LPGNet(Frame)} & emotion2vec & BERT & \textbf{83.87} & \textbf{83.68} \\
\textbf{LPGNet(Utterance)} & emotion2vec & BERT & \textbf{87.99} & \textbf{87.96} \\
\bottomrule
\end{tabular}
\end{table}

\subsection{Ablation Study}

\textbf{Effectiveness of LPIA Components and Modal Interactions.}
To evaluate the role of each component within the LPIA module, we perform a series of ablation studies, as summarized in Table~\ref{tab:ablation1}.

We first examine the impact of removing the intra-modal and inter-modal attention blocks. The results show that eliminating either block leads to a clear performance drop. Specifically, removing inter-modal attention causes a 1.77\% decrease in accuracy, indicating that modeling cross-modal emotional dependencies is crucial for effective fusion. On the other hand, removing intra-modal attention results in a 0.96\% drop, showing that capturing temporal and contextual patterns within each modality also contributes meaningfully to performance. Together, these findings suggest that the two attention types play complementary roles—with intra-modal attention focusing on modality-specific context, and inter-modal attention capturing relationships across modalities.

In addition, we analyze the effect of removing the position-wise feedforward network (FFN), which follows the attention blocks. Although FFN does not directly model alignment, it applies non-linear transformations that refine and enrich the learned representations. Removing this component leads to a 1.12\% drop in F1-score, confirming that the FFN is essential for final feature integration and improved discriminative power.

These results collectively validate the design of the LPIA module, showing that each sub-component contributes to the model’s ability to understand and fuse multimodal emotional cues effectively.

\textbf{Effect of Dual-Gated Fusion.}
To further evaluate the effectiveness of our proposed fusion strategy, we conduct an ablation study by removing the Dual-Gated Fusion module from the model. As reported in Table~\ref{tab:ablation1}, this modification leads to a performance drop in F1-score from 87.96\% to 86.59\%. This confirms the positive impact of our gating mechanism on multimodal fusion.

The Dual-Gated Fusion module contains two key components: a unimodal-level gate and a multimodal-level gate. The unimodal gate filters out irrelevant or noisy features within each modality before fusion, helping the model focus on emotionally relevant content. The multimodal gate then adaptively assigns importance to each modality during the fusion process, allowing the model to balance their contributions based on context.

When this module is removed, the model loses its ability to regulate feature contributions both before and after fusion. This leads to less informative and more noisy multimodal representations. In contrast, the full LPGNet design benefits from dynamic feature calibration at both levels, which is especially important when modalities vary in quality or signal strength. These results demonstrate that combining attention-based interaction modeling (through LPIA) with gated control mechanisms leads to stronger and more robust emotional understanding. The two-stage gating design plays a crucial role in enhancing representation quality, particularly under real-world conditions involving modality imbalance or noise.

\begin{table}[t!]
\centering
\caption{Ablation results of LPGNet.}
\label{tab:ablation1}
\begin{tabular}{lcc}
\toprule
 & ACC (\%) & F1 (\%) \\
\midrule
\textbf{LPGNet} & \textbf{87.99} & \textbf{87.96} \\
\midrule
Modality\\
Text & 81.39 & 81.19 \\
Audio & 84.53 & 84.55 \\
\midrule
LPIA Block\\
w/o inter attention & 86.22 & 86.26 \\
w/o intra attention & 87.03 & 86.95 \\
w/o position-wise FFN & 86.87 & 86.75 \\
\midrule
w/o dual-gated fushion & 86.62 & 86.59 \\
\bottomrule
\end{tabular}
\end{table}

\textbf{Effectiveness of Feature Representation and Multimodal Fusion.} To assess the quality of unimodal feature representations in LPGNet, we compare its performance with existing unimodal baselines. As shown in Table~\ref{tab:audio_comparison} and Table~\ref{tab:text_comparison}, our model consistently outperforms other text-only and audio-only approaches when using the same feature extractors—Emotion2vec for audio and BERT for text. These results confirm that our chosen encoders are effective, and that LPGNet is able to fully leverage them through its attention-based and fusion-oriented architecture.

Importantly, LPGNet maintains strong performance even in unimodal settings. This demonstrates that its architecture can enhance emotion-relevant signals and suppress irrelevant noise, even when limited to a single modality. Compared with simpler baselines such as linear classifiers or prior multimodal models adapted for unimodal input, LPGNet achieves noticeable accuracy gains. This suggests that the model’s structural advantages go beyond feature quality and contribute directly to its discriminative power.

Furthermore, Table~\ref{tab:ablation1} illustrates the added value of integrating multiple modalities. Multimodal fusion significantly improves performance over each unimodal variant, indicating strong cross-modal complementarity. Acoustic and textual modalities contribute different but synergistic cues—such as prosody and semantics—which, when fused, offer a more holistic emotional representation. These findings highlight the necessity of both effective unimodal encoders and carefully designed multimodal integration mechanisms for robust emotion recognition in conversation.

\begin{table}[t!]
\centering
\caption{Comparison with prior unimodal (audio-only) methods.}
\label{tab:audio_comparison}
\begin{tabular}{lcc}
\toprule
Audio Model & Downstream & ACC (\%) \\
\midrule
Wav2Vec & EmotionNAS\cite{sun2022emotionnas} & 69.1 \\
Hubert & MSTR\cite{li2024multi} & 71.6 \\
emotion2vec & Linear\cite{ma2023emotion2vec} & 74.68 \\
emotion2vec & CFIA-Net\cite{hu2024cross} & 76.42 \\
\textbf{emotion2vec} & \textbf{LPGNet} & \textbf{84.53} \\
\bottomrule
\end{tabular}
\end{table}

\begin{table}[t!]
\centering
\caption{Comparison with prior unimodal (text-only) methods.}
\label{tab:text_comparison}
\begin{tabular}{lcc}
\toprule
Text Model & Downstream & ACC (\%) \\
\midrule
bert & MER-HAN\cite{zhang2023multimodal} & 69.54 \\
bert & CwGHP\cite{chauhan2024multimodal} & 79.01 \\
bert & MemoCMT\cite{khan2025memocmt} & 80.66 \\
\textbf{bert (Ours)} & \textbf{LPGNet} & \textbf{81.39} \\
\bottomrule
\end{tabular}
\end{table}

\textbf{Effect of Self-Distillation Coefficients.}
To assess the quality of unimodal feature representations in LPGNet, we compare its performance with existing unimodal baselines. As shown in Table~\ref{tab:audio_comparison} and Table~\ref{tab:text_comparison}, our model consistently outperforms other text-only and audio-only approaches when using the same feature extractor, Emotion2vec for audio and BERT for text. These results confirm that our chosen encoders are effective and that LPGNet can leverage them through its attention-based and fusion-oriented architecture.

Meanwhile, we observe that a lower KL-divergence weight $\lambda_{KL}=0.3$ yields better generalization. Since LPGNet removes speaker embeddings and focuses on speaker-independent modeling, the soft label distributions may exhibit greater inter-sample variability. Down-weighting KL divergence mitigates the risk of overfitting to potentially noisy or over-smoothed soft labels. On the other hand, keeping $\lambda_{CE}=1.0$ ensures that unimodal branches receive sufficient intermediate supervision during distillation.

\begin{table}[t!]
\centering
\caption{Effect of distillation loss weights on performance.}
\label{tab:distillation_weights}
\begin{tabular}{ccc|cc}
\toprule
$\lambda_{task}$ & $\lambda_{CE}$ & $\lambda_{KL}$ & ACC (\%) & F1 (\%) \\
\midrule
1.0 & 1.0 & 1.0 & 87.35 & 87.38 \\
1.0 & 0.0 & 1.0 & 86.36 & 86.53 \\
1.0 & 1.0 & 0.0 & 87.27 & 87.26 \\
1.5 & 1.0 & 0.0 & 87.59 & 87.58 \\
\textbf{1.5} & \textbf{1.0} & \textbf{0.3} & \textbf{87.99} & \textbf{87.96} \\
\bottomrule
\end{tabular}
\end{table}

\begin{table}[t!]
\centering
\caption{Performance comparison with SDT (4-class and 6-class). SDT* uses our feature extractors.}
\label{tab:sdt_comparison}
% \resizebox{0.9\columnwidth}{!}{%
\begin{tabular}{lccccccc}
\toprule
 & Happy & Sad & Neutral & Angry & Excited & Frustrated & Overall \\
\midrule
\textbf{LPGNet (4-class)} & \textbf{92.06} & 85.38 & \textbf{83.99} & 86.75 & -- & -- & \textbf{87.83} \\
\textbf{SDT* (4-class)} & 89.72 & \textbf{86.35} & 82.32 & \textbf{88.15} & -- & -- & 86.62 \\
\midrule
\textbf{LPGNet (6-class)} & \textbf{51.3} & \textbf{79.1} & \textbf{73.99} & 72.19 & \textbf{77.83} & \textbf{68.59} & \textbf{71.78} \\
\textbf{SDT* (6-class)} & 38.59 & 78.09 & 69.52 & \textbf{73.53} & 70.63 & 65.13 & 67.47 \\
\bottomrule
\end{tabular}
% }
\end{table}

\backmatter

\section*{Declarations}

% Some journals require declarations to be submitted in a standardised format. Please check the Instructions for Authors of the journal to which you are submitting to see if you need to complete this section. If yes, your manuscript must contain the following sections under the heading `Declarations':

\begin{itemize}
\item Funding:  This research received no external funding.
\item Conflict of interest/Competing interests: The authors have no relevant financial or non-financial interests to disclose. The authors have no competing interests to declare that are relevant to the content of this article. All authors certify that they have no affiliations with or involvement in any organization or entity with any financial interest or non-financial interest in the subject matter or materials discussed in this manuscript. The authors have no financial or proprietary interests in any material discussed in this article.
\item Ethics approval and consent to participate: Not applicable.
\item Consent for publication: Not applicable.
\item Data availability: The data supporting the findings of this study are publicly available.
\item Materials availability: Not applicable.
\item Code availability: The implementation code for LPGNet will be made available at GitHub publication.
\item Author contribution: Conceptualization, methodology, software and optimization, validation, formal analysis, investigation, writing—original draft preparation,  Zhining He; supervision, writing—review and editing, administration, Yang Xiao.
\end{itemize}

\bibliography{sn-bibliography}% common bib file


\begin{thebibliography}{48}
\ifx \bisbn   \undefined \def \bisbn  #1{ISBN #1}\fi
\ifx \binits  \undefined \def \binits#1{#1}\fi
\ifx \bauthor  \undefined \def \bauthor#1{#1}\fi
\ifx \batitle  \undefined \def \batitle#1{#1}\fi
\ifx \bjtitle  \undefined \def \bjtitle#1{#1}\fi
\ifx \bvolume  \undefined \def \bvolume#1{\textbf{#1}}\fi
\ifx \byear  \undefined \def \byear#1{#1}\fi
\ifx \bissue  \undefined \def \bissue#1{#1}\fi
\ifx \bfpage  \undefined \def \bfpage#1{#1}\fi
\ifx \blpage  \undefined \def \blpage #1{#1}\fi
\ifx \burl  \undefined \def \burl#1{\textsf{#1}}\fi
\ifx \doiurl  \undefined \def \doiurl#1{\url{https://doi.org/#1}}\fi
\ifx \betal  \undefined \def \betal{\textit{et al.}}\fi
\ifx \binstitute  \undefined \def \binstitute#1{#1}\fi
\ifx \binstitutionaled  \undefined \def \binstitutionaled#1{#1}\fi
\ifx \bctitle  \undefined \def \bctitle#1{#1}\fi
\ifx \beditor  \undefined \def \beditor#1{#1}\fi
\ifx \bpublisher  \undefined \def \bpublisher#1{#1}\fi
\ifx \bbtitle  \undefined \def \bbtitle#1{#1}\fi
\ifx \bedition  \undefined \def \bedition#1{#1}\fi
\ifx \bseriesno  \undefined \def \bseriesno#1{#1}\fi
\ifx \blocation  \undefined \def \blocation#1{#1}\fi
\ifx \bsertitle  \undefined \def \bsertitle#1{#1}\fi
\ifx \bsnm \undefined \def \bsnm#1{#1}\fi
\ifx \bsuffix \undefined \def \bsuffix#1{#1}\fi
\ifx \bparticle \undefined \def \bparticle#1{#1}\fi
\ifx \barticle \undefined \def \barticle#1{#1}\fi
\bibcommenthead
\ifx \bconfdate \undefined \def \bconfdate #1{#1}\fi
\ifx \botherref \undefined \def \botherref #1{#1}\fi
\ifx \url \undefined \def \url#1{\textsf{#1}}\fi
\ifx \bchapter \undefined \def \bchapter#1{#1}\fi
\ifx \bbook \undefined \def \bbook#1{#1}\fi
\ifx \bcomment \undefined \def \bcomment#1{#1}\fi
\ifx \oauthor \undefined \def \oauthor#1{#1}\fi
\ifx \citeauthoryear \undefined \def \citeauthoryear#1{#1}\fi
\ifx \endbibitem  \undefined \def \endbibitem {}\fi
\ifx \bconflocation  \undefined \def \bconflocation#1{#1}\fi
\ifx \arxivurl  \undefined \def \arxivurl#1{\textsf{#1}}\fi
\csname PreBibitemsHook\endcsname

%%% 1
\bibitem[\protect\citeauthoryear{Poria et~al.}{2019}]{poria2019emotion}
\begin{barticle}
\bauthor{\bsnm{Poria}, \binits{S.}},
\bauthor{\bsnm{Majumder}, \binits{N.}},
\bauthor{\bsnm{Mihalcea}, \binits{R.}},
\bauthor{\bsnm{Hovy}, \binits{E.}}:
\batitle{Emotion recognition in conversation: Research challenges, datasets, and recent advances}.
\bjtitle{IEEE access}
\bvolume{7},
\bfpage{100943}--\blpage{100953}
(\byear{2019})
\end{barticle}
\endbibitem

%%% 2
\bibitem[\protect\citeauthoryear{Qin et~al.}{2023}]{qin2023bert}
\begin{bchapter}
\bauthor{\bsnm{Qin}, \binits{X.}},
\bauthor{\bsnm{Wu}, \binits{Z.}},
\bauthor{\bsnm{Zhang}, \binits{T.}},
\bauthor{\bsnm{Li}, \binits{Y.}},
\bauthor{\bsnm{Luan}, \binits{J.}},
\bauthor{\bsnm{Wang}, \binits{B.}},
\bauthor{\bsnm{Wang}, \binits{L.}},
\bauthor{\bsnm{Cui}, \binits{J.}}:
\bctitle{Bert-erc: Fine-tuning bert is enough for emotion recognition in conversation}.
In: \bbtitle{Proceedings of the AAAI Conference on Artificial Intelligence},
vol. \bseriesno{37},
pp. \bfpage{13492}--\blpage{13500}
(\byear{2023})
\end{bchapter}
\endbibitem

%%% 3
\bibitem[\protect\citeauthoryear{Ramakrishnan and El~Emary}{2013}]{ramakrishnan2013speech}
\begin{barticle}
\bauthor{\bsnm{Ramakrishnan}, \binits{S.}},
\bauthor{\bsnm{El~Emary}, \binits{I.M.}}:
\batitle{Speech emotion recognition approaches in human computer interaction}.
\bjtitle{Telecommunication Systems}
\bvolume{52},
\bfpage{1467}--\blpage{1478}
(\byear{2013})
\end{barticle}
\endbibitem

%%% 4
\bibitem[\protect\citeauthoryear{Cowie et~al.}{2001}]{cowie2001emotion}
\begin{barticle}
\bauthor{\bsnm{Cowie}, \binits{R.}},
\bauthor{\bsnm{Douglas-Cowie}, \binits{E.}},
\bauthor{\bsnm{Tsapatsoulis}, \binits{N.}},
\bauthor{\bsnm{Votsis}, \binits{G.}},
\bauthor{\bsnm{Kollias}, \binits{S.}},
\bauthor{\bsnm{Fellenz}, \binits{W.}},
\bauthor{\bsnm{Taylor}, \binits{J.G.}}:
\batitle{Emotion recognition in human-computer interaction}.
\bjtitle{IEEE Signal processing magazine}
\bvolume{18}(\bissue{1}),
\bfpage{32}--\blpage{80}
(\byear{2001})
\end{barticle}
\endbibitem

%%% 5
\bibitem[\protect\citeauthoryear{Ma et~al.}{2023}]{ma2023transformer}
\begin{barticle}
\bauthor{\bsnm{Ma}, \binits{H.}},
\bauthor{\bsnm{Wang}, \binits{J.}},
\bauthor{\bsnm{Lin}, \binits{H.}},
\bauthor{\bsnm{Zhang}, \binits{B.}},
\bauthor{\bsnm{Zhang}, \binits{Y.}},
\bauthor{\bsnm{Xu}, \binits{B.}}:
\batitle{A transformer-based model with self-distillation for multimodal emotion recognition in conversations}.
\bjtitle{IEEE Transactions on Multimedia}
\bvolume{26},
\bfpage{776}--\blpage{788}
(\byear{2023})
\end{barticle}
\endbibitem

%%% 6
\bibitem[\protect\citeauthoryear{Ghosal et~al.}{2019}]{ghosal2019dialoguegcn}
\begin{botherref}
\oauthor{\bsnm{Ghosal}, \binits{D.}},
\oauthor{\bsnm{Majumder}, \binits{N.}},
\oauthor{\bsnm{Poria}, \binits{S.}},
\oauthor{\bsnm{Chhaya}, \binits{N.}},
\oauthor{\bsnm{Gelbukh}, \binits{A.}}:
Dialoguegcn: A graph convolutional neural network for emotion recognition in conversation.
arXiv preprint arXiv:1908.11540
(2019)
\end{botherref}
\endbibitem

%%% 7
\bibitem[\protect\citeauthoryear{Chen et~al.}{2023}]{chen2023multivariate}
\begin{bchapter}
\bauthor{\bsnm{Chen}, \binits{F.}},
\bauthor{\bsnm{Shao}, \binits{J.}},
\bauthor{\bsnm{Zhu}, \binits{S.}},
\bauthor{\bsnm{Shen}, \binits{H.T.}}:
\bctitle{Multivariate, multi-frequency and multimodal: Rethinking graph neural networks for emotion recognition in conversation}.
In: \bbtitle{Proceedings of the IEEE/CVF Conference on Computer Vision and Pattern Recognition},
pp. \bfpage{10761}--\blpage{10770}
(\byear{2023})
\end{bchapter}
\endbibitem

%%% 8
\bibitem[\protect\citeauthoryear{Song et~al.}{2023}]{song2023sunet}
\begin{barticle}
\bauthor{\bsnm{Song}, \binits{R.}},
\bauthor{\bsnm{Giunchiglia}, \binits{F.}},
\bauthor{\bsnm{Shi}, \binits{L.}},
\bauthor{\bsnm{Shen}, \binits{Q.}},
\bauthor{\bsnm{Xu}, \binits{H.}}:
\batitle{Sunet: Speaker-utterance interaction graph neural network for emotion recognition in conversations}.
\bjtitle{Engineering Applications of Artificial Intelligence}
\bvolume{123},
\bfpage{106315}
(\byear{2023})
\end{barticle}
\endbibitem

%%% 9
\bibitem[\protect\citeauthoryear{Tang et~al.}{2022}]{tang2022bimodal}
\begin{barticle}
\bauthor{\bsnm{Tang}, \binits{Y.}},
\bauthor{\bsnm{Hu}, \binits{Y.}},
\bauthor{\bsnm{He}, \binits{L.}},
\bauthor{\bsnm{Huang}, \binits{H.}}:
\batitle{A bimodal network based on audio--text-interactional-attention with arcface loss for speech emotion recognition}.
\bjtitle{Speech Communication}
\bvolume{143},
\bfpage{21}--\blpage{32}
(\byear{2022})
\end{barticle}
\endbibitem

%%% 10
\bibitem[\protect\citeauthoryear{Zhao and Curtis}{2023}]{zhao2023bayesian}
\begin{botherref}
\oauthor{\bsnm{Zhao}, \binits{X.}},
\oauthor{\bsnm{Curtis}, \binits{A.}}:
Bayesian inversion, uncertainty analysis and interrogation using boosting variational inference.
arXiv preprint arXiv:2312.17646
(2023)
\end{botherref}
\endbibitem

%%% 11
\bibitem[\protect\citeauthoryear{Maji et~al.}{2023}]{maji2023multimodal}
\begin{bchapter}
\bauthor{\bsnm{Maji}, \binits{B.}},
\bauthor{\bsnm{Swain}, \binits{M.}},
\bauthor{\bsnm{Guha}, \binits{R.}},
\bauthor{\bsnm{Routray}, \binits{A.}}:
\bctitle{Multimodal emotion recognition based on deep temporal features using cross-modal transformer and self-attention}.
In: \bbtitle{ICASSP 2023-2023 IEEE International Conference on Acoustics, Speech and Signal Processing (ICASSP)},
pp. \bfpage{1}--\blpage{5}
(\byear{2023}).
\bcomment{IEEE}
\end{bchapter}
\endbibitem

%%% 12
\bibitem[\protect\citeauthoryear{Hu et~al.}{2021}]{hu2021mmgcn}
\begin{botherref}
\oauthor{\bsnm{Hu}, \binits{J.}},
\oauthor{\bsnm{Liu}, \binits{Y.}},
\oauthor{\bsnm{Zhao}, \binits{J.}},
\oauthor{\bsnm{Jin}, \binits{Q.}}:
Mmgcn: Multimodal fusion via deep graph convolution network for emotion recognition in conversation.
arXiv preprint arXiv:2107.06779
(2021)
\end{botherref}
\endbibitem

%%% 13
\bibitem[\protect\citeauthoryear{Peng and Xiao}{2025}]{cl1}
\begin{bchapter}
\bauthor{\bsnm{Peng}, \binits{T.}},
\bauthor{\bsnm{Xiao}, \binits{Y.}}:
\bctitle{Dark experience for incremental keyword spotting}.
In: \bbtitle{ICASSP 2025-2025 IEEE International Conference on Acoustics, Speech and Signal Processing (ICASSP)},
pp. \bfpage{1}--\blpage{5}
(\byear{2025}).
\bcomment{IEEE}
\end{bchapter}
\endbibitem

%%% 14
\bibitem[\protect\citeauthoryear{Xiao et~al.}{2025}]{cl2}
\begin{botherref}
\oauthor{\bsnm{Xiao}, \binits{Y.}},
\oauthor{\bsnm{Peng}, \binits{T.}},
\oauthor{\bsnm{Das}, \binits{R.K.}},
\oauthor{\bsnm{Hu}, \binits{Y.}},
\oauthor{\bsnm{Zhuang}, \binits{H.}}:
Analytickws: towards exemplar-free analytic class incremental learning for small-footprint keyword spotting.
arXiv preprint:2505.11817
(2025)
\end{botherref}
\endbibitem

%%% 15
\bibitem[\protect\citeauthoryear{Xiao et~al.}{2024}]{cl3}
\begin{botherref}
\oauthor{\bsnm{Xiao}, \binits{Y.}},
\oauthor{\bsnm{Yin}, \binits{H.}},
\oauthor{\bsnm{Bai}, \binits{J.}},
\oauthor{\bsnm{Das}, \binits{R.K.}}:
Mixstyle based domain generalization for sound event detection with heterogeneous training data.
arXiv preprint:2407.03654
(2024)
\end{botherref}
\endbibitem

%%% 16
\bibitem[\protect\citeauthoryear{Xiao and Das}{2024}]{ssl1}
\begin{botherref}
\oauthor{\bsnm{Xiao}, \binits{Y.}},
\oauthor{\bsnm{Das}, \binits{R.K.}}:
Where's that voice coming? continual learning for sound source localization.
arXiv preprint:2407.03661
(2024)
\end{botherref}
\endbibitem

%%% 17
\bibitem[\protect\citeauthoryear{Chang et~al.}{2024}]{llm1}
\begin{barticle}
\bauthor{\bsnm{Chang}, \binits{Y.}},
\bauthor{\bsnm{Wang}, \binits{X.}},
\bauthor{\bsnm{Wang}, \binits{J.}},
\bauthor{\bsnm{Wu}, \binits{Y.}},
\bauthor{\bsnm{Yang}, \binits{L.}},
\bauthor{\bsnm{Zhu}, \binits{K.}},
\bauthor{\bsnm{Chen}, \binits{H.}},
\bauthor{\bsnm{Yi}, \binits{X.}},
\bauthor{\bsnm{Wang}, \binits{C.}},
\bauthor{\bsnm{Wang}, \binits{Y.}}, \betal:
\batitle{A survey on evaluation of large language models}.
\bjtitle{ACM transactions on intelligent systems and technology}
\bvolume{15},
\bfpage{1}--\blpage{45}
(\byear{2024})
\end{barticle}
\endbibitem

%%% 18
\bibitem[\protect\citeauthoryear{Naveed et~al.}{2023}]{llm2}
\begin{botherref}
\oauthor{\bsnm{Naveed}, \binits{H.}},
\oauthor{\bsnm{Khan}, \binits{A.U.}},
\oauthor{\bsnm{Qiu}, \binits{S.}},
\oauthor{\bsnm{Saqib}, \binits{M.}},
\oauthor{\bsnm{Anwar}, \binits{S.}},
\oauthor{\bsnm{Usman}, \binits{M.}},
\oauthor{\bsnm{Akhtar}, \binits{N.}},
\oauthor{\bsnm{Barnes}, \binits{N.}},
\oauthor{\bsnm{Mian}, \binits{A.}}:
A comprehensive overview of large language models.
ACM Transactions on Intelligent Systems and Technology
(2023)
\end{botherref}
\endbibitem

%%% 19
\bibitem[\protect\citeauthoryear{Xiao and Das}{2024}]{llm3}
\begin{botherref}
\oauthor{\bsnm{Xiao}, \binits{Y.}},
\oauthor{\bsnm{Das}, \binits{R.K.}}:
Wilddesed: an llm-powered dataset for wild domestic environment sound event detection system.
arXiv preprint:2407.03656
(2024)
\end{botherref}
\endbibitem

%%% 20
\bibitem[\protect\citeauthoryear{Wu et~al.}{2024}]{wu2024beyond}
\begin{botherref}
\oauthor{\bsnm{Wu}, \binits{Z.}},
\oauthor{\bsnm{Gong}, \binits{Z.}},
\oauthor{\bsnm{Ai}, \binits{L.}},
\oauthor{\bsnm{Shi}, \binits{P.}},
\oauthor{\bsnm{Donbekci}, \binits{K.}},
\oauthor{\bsnm{Hirschberg}, \binits{J.}}:
Beyond silent letters: Amplifying llms in emotion recognition with vocal nuances.
arXiv preprint arXiv:2407.21315
(2024)
\end{botherref}
\endbibitem

%%% 21
\bibitem[\protect\citeauthoryear{Zhang et~al.}{2025}]{zhang2025dialoguellm}
\begin{botherref}
\oauthor{\bsnm{Zhang}, \binits{Y.}},
\oauthor{\bsnm{Wang}, \binits{M.}},
\oauthor{\bsnm{Wu}, \binits{Y.}},
\oauthor{\bsnm{Tiwari}, \binits{P.}},
\oauthor{\bsnm{Li}, \binits{Q.}},
\oauthor{\bsnm{Wang}, \binits{B.}},
\oauthor{\bsnm{Qin}, \binits{J.}}:
Dialoguellm: Context and emotion knowledge-tuned large language models for emotion recognition in conversations.
Neural Networks,
107901
(2025)
\end{botherref}
\endbibitem

%%% 22
\bibitem[\protect\citeauthoryear{Xue et~al.}{2024}]{xue2024bioserc}
\begin{bchapter}
\bauthor{\bsnm{Xue}, \binits{J.}},
\bauthor{\bsnm{Nguyen}, \binits{M.-P.}},
\bauthor{\bsnm{Matheny}, \binits{B.}},
\bauthor{\bsnm{Nguyen}, \binits{L.-M.}}:
\bctitle{Bioserc: Integrating biography speakers supported by llms for erc tasks}.
In: \bbtitle{International Conference on Artificial Neural Networks},
pp. \bfpage{277}--\blpage{292}
(\byear{2024}).
\bcomment{Springer}
\end{bchapter}
\endbibitem

%%% 23
\bibitem[\protect\citeauthoryear{Zhang et~al.}{2019}]{zhang2019your}
\begin{bchapter}
\bauthor{\bsnm{Zhang}, \binits{L.}},
\bauthor{\bsnm{Song}, \binits{J.}},
\bauthor{\bsnm{Gao}, \binits{A.}},
\bauthor{\bsnm{Chen}, \binits{J.}},
\bauthor{\bsnm{Bao}, \binits{C.}},
\bauthor{\bsnm{Ma}, \binits{K.}}:
\bctitle{Be your own teacher: Improve the performance of convolutional neural networks via self distillation}.
In: \bbtitle{Proceedings of the IEEE/CVF International Conference on Computer Vision},
pp. \bfpage{3713}--\blpage{3722}
(\byear{2019})
\end{bchapter}
\endbibitem

%%% 24
\bibitem[\protect\citeauthoryear{Zhang et~al.}{2021}]{zhang2021self}
\begin{barticle}
\bauthor{\bsnm{Zhang}, \binits{L.}},
\bauthor{\bsnm{Bao}, \binits{C.}},
\bauthor{\bsnm{Ma}, \binits{K.}}:
\batitle{Self-distillation: Towards efficient and compact neural networks}.
\bjtitle{IEEE Transactions on Pattern Analysis and Machine Intelligence}
\bvolume{44}(\bissue{8}),
\bfpage{4388}--\blpage{4403}
(\byear{2021})
\end{barticle}
\endbibitem

%%% 25
\bibitem[\protect\citeauthoryear{Xiao and Das}{2024}]{xiao2024ucil}
\begin{botherref}
\oauthor{\bsnm{Xiao}, \binits{Y.}},
\oauthor{\bsnm{Das}, \binits{R.K.}}:
Ucil: An unsupervised class incremental learning approach for sound event detection.
arXiv preprint:2407.03657
(2024)
\end{botherref}
\endbibitem

%%% 26
\bibitem[\protect\citeauthoryear{Li et~al.}{2022}]{li2022teacher}
\begin{bchapter}
\bauthor{\bsnm{Li}, \binits{L.}},
\bauthor{\bsnm{Liang}, \binits{S.-N.}},
\bauthor{\bsnm{Yang}, \binits{Y.}},
\bauthor{\bsnm{Jin}, \binits{Z.}}:
\bctitle{Teacher-free distillation via regularizing intermediate representation}.
In: \bbtitle{2022 International Joint Conference on Neural Networks (IJCNN)},
pp. \bfpage{01}--\blpage{06}
(\byear{2022}).
\bcomment{IEEE}
\end{bchapter}
\endbibitem

%%% 27
\bibitem[\protect\citeauthoryear{Zhang et~al.}{2024}]{zhang2024deep}
\begin{barticle}
\bauthor{\bsnm{Zhang}, \binits{S.}},
\bauthor{\bsnm{Yang}, \binits{Y.}},
\bauthor{\bsnm{Chen}, \binits{C.}},
\bauthor{\bsnm{Zhang}, \binits{X.}},
\bauthor{\bsnm{Leng}, \binits{Q.}},
\bauthor{\bsnm{Zhao}, \binits{X.}}:
\batitle{Deep learning-based multimodal emotion recognition from audio, visual, and text modalities: A systematic review of recent advancements and future prospects}.
\bjtitle{Expert Systems with Applications}
\bvolume{237},
\bfpage{121692}
(\byear{2024})
\end{barticle}
\endbibitem

%%% 28
\bibitem[\protect\citeauthoryear{Hazarika et~al.}{2018a}]{hazarika2018conversational}
\begin{bchapter}
\bauthor{\bsnm{Hazarika}, \binits{D.}},
\bauthor{\bsnm{Poria}, \binits{S.}},
\bauthor{\bsnm{Zadeh}, \binits{A.}},
\bauthor{\bsnm{Cambria}, \binits{E.}},
\bauthor{\bsnm{Morency}, \binits{L.-P.}},
\bauthor{\bsnm{Zimmermann}, \binits{R.}}:
\bctitle{Conversational memory network for emotion recognition in dyadic dialogue videos}.
In: \bbtitle{Proceedings of the Conference. Association for Computational Linguistics. North American Chapter. Meeting},
vol. \bseriesno{2018},
p. \bfpage{2122}
(\byear{2018})
\end{bchapter}
\endbibitem

%%% 29
\bibitem[\protect\citeauthoryear{Hazarika et~al.}{2018b}]{hazarika2018icon}
\begin{bchapter}
\bauthor{\bsnm{Hazarika}, \binits{D.}},
\bauthor{\bsnm{Poria}, \binits{S.}},
\bauthor{\bsnm{Mihalcea}, \binits{R.}},
\bauthor{\bsnm{Cambria}, \binits{E.}},
\bauthor{\bsnm{Zimmermann}, \binits{R.}}:
\bctitle{Icon: Interactive conversational memory network for multimodal emotion detection}.
In: \bbtitle{Proceedings of the 2018 Conference on Empirical Methods in Natural Language Processing},
pp. \bfpage{2594}--\blpage{2604}
(\byear{2018})
\end{bchapter}
\endbibitem

%%% 30
\bibitem[\protect\citeauthoryear{Zhang et~al.}{2023}]{zhang2023multimodal}
\begin{barticle}
\bauthor{\bsnm{Zhang}, \binits{S.}},
\bauthor{\bsnm{Yang}, \binits{Y.}},
\bauthor{\bsnm{Chen}, \binits{C.}},
\bauthor{\bsnm{Liu}, \binits{R.}},
\bauthor{\bsnm{Tao}, \binits{X.}},
\bauthor{\bsnm{Guo}, \binits{W.}},
\bauthor{\bsnm{Xu}, \binits{Y.}},
\bauthor{\bsnm{Zhao}, \binits{X.}}:
\batitle{Multimodal emotion recognition based on audio and text by using hybrid attention networks}.
\bjtitle{Biomedical Signal Processing and Control}
\bvolume{85},
\bfpage{105052}
(\byear{2023})
\end{barticle}
\endbibitem

%%% 31
\bibitem[\protect\citeauthoryear{Nguyen et~al.}{2025}]{nguyen2025enhanced}
\begin{barticle}
\bauthor{\bsnm{Nguyen}, \binits{D.-K.}},
\bauthor{\bsnm{Lim}, \binits{E.}},
\bauthor{\bsnm{Kim}, \binits{S.-H.}},
\bauthor{\bsnm{Yang}, \binits{H.-J.}},
\bauthor{\bsnm{Kim}, \binits{S.}}:
\batitle{Enhanced emotion recognition through dynamic restrained adaptive loss and extended multimodal bottleneck transformer}.
\bjtitle{Applied Sciences}
\bvolume{15}(\bissue{5}),
\bfpage{2862}
(\byear{2025})
\end{barticle}
\endbibitem

%%% 32
\bibitem[\protect\citeauthoryear{Ghosal et~al.}{2020}]{ghosal2020cosmic}
\begin{botherref}
\oauthor{\bsnm{Ghosal}, \binits{D.}},
\oauthor{\bsnm{Majumder}, \binits{N.}},
\oauthor{\bsnm{Gelbukh}, \binits{A.}},
\oauthor{\bsnm{Mihalcea}, \binits{R.}},
\oauthor{\bsnm{Poria}, \binits{S.}}:
Cosmic: Commonsense knowledge for emotion identification in conversations.
arXiv preprint arXiv:2010.02795
(2020)
\end{botherref}
\endbibitem

%%% 33
\bibitem[\protect\citeauthoryear{Shen et~al.}{2021}]{shen2021dialogxl}
\begin{bchapter}
\bauthor{\bsnm{Shen}, \binits{W.}},
\bauthor{\bsnm{Chen}, \binits{J.}},
\bauthor{\bsnm{Quan}, \binits{X.}},
\bauthor{\bsnm{Xie}, \binits{Z.}}:
\bctitle{Dialogxl: All-in-one xlnet for multi-party conversation emotion recognition}.
In: \bbtitle{Proceedings of the AAAI Conference on Artificial Intelligence},
vol. \bseriesno{35},
pp. \bfpage{13789}--\blpage{13797}
(\byear{2021})
\end{bchapter}
\endbibitem

%%% 34
\bibitem[\protect\citeauthoryear{Kuhlen and Rahman}{2022}]{kuhlen2022mental}
\begin{barticle}
\bauthor{\bsnm{Kuhlen}, \binits{A.K.}},
\bauthor{\bsnm{Rahman}, \binits{R.A.}}:
\batitle{Mental chronometry of speaking in dialogue: Semantic interference turns into facilitation}.
\bjtitle{Cognition}
\bvolume{219},
\bfpage{104962}
(\byear{2022})
\end{barticle}
\endbibitem

%%% 35
\bibitem[\protect\citeauthoryear{Yin et~al.}{2022}]{yin2022mix}
\begin{bchapter}
\bauthor{\bsnm{Yin}, \binits{Y.}},
\bauthor{\bsnm{Zhu}, \binits{B.}},
\bauthor{\bsnm{Chen}, \binits{J.}},
\bauthor{\bsnm{Cheng}, \binits{L.}},
\bauthor{\bsnm{Jiang}, \binits{Y.-G.}}:
\bctitle{Mix-dann and dynamic-modal-distillation for video domain adaptation}.
In: \bbtitle{Proceedings of the 30th ACM International Conference on Multimedia},
pp. \bfpage{3224}--\blpage{3233}
(\byear{2022})
\end{bchapter}
\endbibitem

%%% 36
\bibitem[\protect\citeauthoryear{Aslam et~al.}{2024}]{aslam2024multi}
\begin{botherref}
\oauthor{\bsnm{Aslam}, \binits{M.H.}},
\oauthor{\bsnm{Pedersoli}, \binits{M.}},
\oauthor{\bsnm{Koerich}, \binits{A.L.}},
\oauthor{\bsnm{Granger}, \binits{E.}}:
Multi teacher privileged knowledge distillation for multimodal expression recognition.
arXiv preprint arXiv:2408.09035
(2024)
\end{botherref}
\endbibitem

%%% 37
\bibitem[\protect\citeauthoryear{Busso et~al.}{2008}]{busso2008iemocap}
\begin{barticle}
\bauthor{\bsnm{Busso}, \binits{C.}},
\bauthor{\bsnm{Bulut}, \binits{M.}},
\bauthor{\bsnm{Lee}, \binits{C.-C.}},
\bauthor{\bsnm{Kazemzadeh}, \binits{A.}},
\bauthor{\bsnm{Mower}, \binits{E.}},
\bauthor{\bsnm{Kim}, \binits{S.}},
\bauthor{\bsnm{Chang}, \binits{J.N.}},
\bauthor{\bsnm{Lee}, \binits{S.}},
\bauthor{\bsnm{Narayanan}, \binits{S.S.}}:
\batitle{Iemocap: Interactive emotional dyadic motion capture database}.
\bjtitle{Language resources and evaluation}
\bvolume{42},
\bfpage{335}--\blpage{359}
(\byear{2008})
\end{barticle}
\endbibitem

%%% 38
\bibitem[\protect\citeauthoryear{Li et~al.}{2025}]{li2025gatedxlstm}
\begin{botherref}
\oauthor{\bsnm{Li}, \binits{Y.}},
\oauthor{\bsnm{Sun}, \binits{Q.}},
\oauthor{\bsnm{Murthy}, \binits{S.M.K.}},
\oauthor{\bsnm{Alturki}, \binits{E.}},
\oauthor{\bsnm{Schuller}, \binits{B.W.}}:
Gatedxlstm: A multimodal affective computing approach for emotion recognition in conversations.
arXiv preprint arXiv:2503.20919
(2025)
\end{botherref}
\endbibitem

%%% 39
\bibitem[\protect\citeauthoryear{Wang et~al.}{2025a}]{wang2025emotion}
\begin{bchapter}
\bauthor{\bsnm{Wang}, \binits{J.}},
\bauthor{\bsnm{Li}, \binits{N.}},
\bauthor{\bsnm{Zhang}, \binits{L.}},
\bauthor{\bsnm{Shan}, \binits{L.}}:
\bctitle{Emotion recognition for multimodal information interaction}.
In: \bbtitle{2025 International Conference on Intelligent Systems and Computational Networks (ICISCN)},
pp. \bfpage{1}--\blpage{7}
(\byear{2025}).
\bcomment{IEEE}
\end{bchapter}
\endbibitem

%%% 40
\bibitem[\protect\citeauthoryear{Wang et~al.}{2025b}]{wang2025enhancing}
\begin{bchapter}
\bauthor{\bsnm{Wang}, \binits{X.}},
\bauthor{\bsnm{Zhao}, \binits{S.}},
\bauthor{\bsnm{Sun}, \binits{H.}},
\bauthor{\bsnm{Wang}, \binits{H.}},
\bauthor{\bsnm{Zhou}, \binits{J.}},
\bauthor{\bsnm{Qin}, \binits{Y.}}:
\bctitle{Enhancing multimodal emotion recognition through multi-granularity cross-modal alignment}.
In: \bbtitle{ICASSP 2025-2025 IEEE International Conference on Acoustics, Speech and Signal Processing (ICASSP)},
pp. \bfpage{1}--\blpage{5}
(\byear{2025}).
\bcomment{IEEE}
\end{bchapter}
\endbibitem

%%% 41
\bibitem[\protect\citeauthoryear{Shi et~al.}{2023}]{shi2023emotion}
\begin{bchapter}
\bauthor{\bsnm{Shi}, \binits{X.}},
\bauthor{\bsnm{Li}, \binits{X.}},
\bauthor{\bsnm{Toda}, \binits{T.}}:
\bctitle{Emotion awareness in multi-utterance turn for improving emotion prediction in multi-speaker conversation}.
In: \bbtitle{Proc. Interspeech},
vol. \bseriesno{2023},
pp. \bfpage{765}--\blpage{769}
(\byear{2023})
\end{bchapter}
\endbibitem

%%% 42
\bibitem[\protect\citeauthoryear{Adeel and Tao}{2024}]{adeel2024enhancing}
\begin{bchapter}
\bauthor{\bsnm{Adeel}, \binits{M.}},
\bauthor{\bsnm{Tao}, \binits{Z.-Y.}}:
\bctitle{Enhancing speech emotion recognition in urdu using bi-gru networks: An in-depth analysis of acoustic features and model interpretability}.
In: \bbtitle{2024 IEEE International Conference on Industrial Technology (ICIT)},
pp. \bfpage{1}--\blpage{6}
(\byear{2024}).
\bcomment{IEEE}
\end{bchapter}
\endbibitem

%%% 43
\bibitem[\protect\citeauthoryear{Khan et~al.}{2025}]{khan2025memocmt}
\begin{barticle}
\bauthor{\bsnm{Khan}, \binits{M.}},
\bauthor{\bsnm{Tran}, \binits{P.-N.}},
\bauthor{\bsnm{Pham}, \binits{N.T.}},
\bauthor{\bsnm{El~Saddik}, \binits{A.}},
\bauthor{\bsnm{Othmani}, \binits{A.}}:
\batitle{Memocmt: multimodal emotion recognition using cross-modal transformer-based feature fusion}.
\bjtitle{Scientific reports}
\bvolume{15}(\bissue{1}),
\bfpage{5473}
(\byear{2025})
\end{barticle}
\endbibitem

%%% 44
\bibitem[\protect\citeauthoryear{Ma et~al.}{2023}]{ma2023emotion2vec}
\begin{botherref}
\oauthor{\bsnm{Ma}, \binits{Z.}},
\oauthor{\bsnm{Zheng}, \binits{Z.}},
\oauthor{\bsnm{Ye}, \binits{J.}},
\oauthor{\bsnm{Li}, \binits{J.}},
\oauthor{\bsnm{Gao}, \binits{Z.}},
\oauthor{\bsnm{Zhang}, \binits{S.}},
\oauthor{\bsnm{Chen}, \binits{X.}}:
emotion2vec: Self-supervised pre-training for speech emotion representation.
arXiv preprint arXiv:2312.15185
(2023)
\end{botherref}
\endbibitem

%%% 45
\bibitem[\protect\citeauthoryear{Hu et~al.}{2024}]{hu2024cross}
\begin{bchapter}
\bauthor{\bsnm{Hu}, \binits{Y.}},
\bauthor{\bsnm{Yang}, \binits{H.}},
\bauthor{\bsnm{Huang}, \binits{H.}},
\bauthor{\bsnm{He}, \binits{L.}}:
\bctitle{Cross-modal features interaction-and-aggregation network with self-consistency training for speech emotion recognition [c]}.
In: \bbtitle{Proc. Interspeech 2024},
pp. \bfpage{2335}--\blpage{2339}
(\byear{2024})
\end{bchapter}
\endbibitem

%%% 46
\bibitem[\protect\citeauthoryear{Sun et~al.}{2022}]{sun2022emotionnas}
\begin{botherref}
\oauthor{\bsnm{Sun}, \binits{H.}},
\oauthor{\bsnm{Lian}, \binits{Z.}},
\oauthor{\bsnm{Liu}, \binits{B.}},
\oauthor{\bsnm{Li}, \binits{Y.}},
\oauthor{\bsnm{Sun}, \binits{L.}},
\oauthor{\bsnm{Cai}, \binits{C.}},
\oauthor{\bsnm{Tao}, \binits{J.}},
\oauthor{\bsnm{Wang}, \binits{M.}},
\oauthor{\bsnm{Cheng}, \binits{Y.}}:
Emotionnas: Two-stream neural architecture search for speech emotion recognition.
arXiv preprint arXiv:2203.13617
(2022)
\end{botherref}
\endbibitem

%%% 47
\bibitem[\protect\citeauthoryear{Li et~al.}{2024}]{li2024multi}
\begin{botherref}
\oauthor{\bsnm{Li}, \binits{Z.}},
\oauthor{\bsnm{Xing}, \binits{X.}},
\oauthor{\bsnm{Fang}, \binits{Y.}},
\oauthor{\bsnm{Zhang}, \binits{W.}},
\oauthor{\bsnm{Fan}, \binits{H.}},
\oauthor{\bsnm{Xu}, \binits{X.}}:
Multi-scale temporal transformer for speech emotion recognition.
arXiv preprint arXiv:2410.00390
(2024)
\end{botherref}
\endbibitem

%%% 48
\bibitem[\protect\citeauthoryear{Chauhan et~al.}{2024}]{chauhan2024multimodal}
\begin{barticle}
\bauthor{\bsnm{Chauhan}, \binits{K.}},
\bauthor{\bsnm{Sharma}, \binits{K.K.}},
\bauthor{\bsnm{Varma}, \binits{T.}}:
\batitle{Multimodal emotion recognition using contextualized audio information and ground transcripts on multiple datasets}.
\bjtitle{Arabian Journal for Science and Engineering}
\bvolume{49}(\bissue{9}),
\bfpage{11871}--\blpage{11881}
(\byear{2024})
\end{barticle}
\endbibitem

\end{thebibliography}

\end{document}